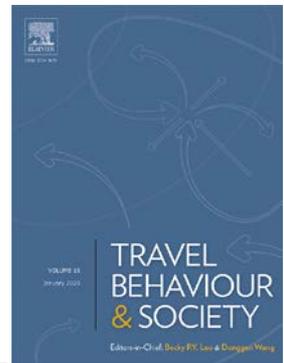

# Latent learning and the formation of a spatiotemporal cognitive map of a road network

Navid Khademi[a,*], Ramin Saedi[b]

[a] *School of Civil Engineering, College of Engineering, University of Tehran, Tehran, Iran.*
[b] *Department of Civil and Environmental Engineering, Michigan State University, East Lansing, MI 48824-1226.*

**Abstract**

This paper discusses two main themes. First, it investigates the formation of a spatiotemporal cognitive map (mental image) of a road network in travelers' memory, which entails the travelers' global conceptual understanding of congestion or the degree of crowding of the network. Second, it tries to investigate how latent learning of travelers from previous experiences shapes parts of the mental image, even for the parts of the network with which the travelers are unfamiliar. An experiment of route choice experiences was conducted among 90 participants in order to gain insight into the formation of a cognitive map and latent learning. In this experiment, the following independent variables are connected to the formation of a mental image of the network and the quality of the generalization of the unfamiliar parts of the network: (i) dispersion of the links' travel time throughout the network, (ii) number of trips the traveler makes, (iii) traveler's gender, (iv) traveler's driving experiences, (v) traveler's natural level of optimism or pessimism, (vi) salient or noticeable features on the network, and (vii) the presence of traffic signals. Several nonparametric (distribution-free) tests are employed to test the hypotheses. The results indicate that apart from the traveler's gender and salient or noticeable features on the network, the considered factors significantly affect the degree of the recognizability of the network elements by travelers.

---

[*] Corresponding author's address: School of Civil Engineering, College of Engineering, University of Tehran, 16–Azar Street, Tehran, Iran. P.O. Box: 1417613131. Phone: +98 (0)9127259280, Fax: +98 (0)21 66403808 e–mail: navid.khademi@ut.ac.ir

## Introduction

Modeling travelers' choice behavior and predicting the reactions to previous experiences is viewed as an essential component of the new era of transportation analysis methods. Travelers usually plan their trips based on their cognition, experiences, external data received from trip information systems, the community, etc. In order to understand how travelers acquire information and how this information is processed and utilized for subsequent decisions, it is necessary to have a deeper understanding of the travelers' information acquisition, persistence of this information in their memory, their ability in recalling this information, and in a general concept, their cognition.

Transportation network analysis methods are anticipated to shift from static models to activity-based and dynamic ones. Activity-based models are a convenient way to derive time-location patterns of people, and to predict characteristics of activities such as timing, duration, location, and people involved (Rasouli and Timmermans, 2014). Daily activity-based trips vary from one day to another, implying a dynamic nature of demand modeling. In this context, travelers' cognitive maps of the transportation network may change on a daily basis. Knowledge of the daily adjustment of the pattern of the travelers' cognitive maps can lead to the development of capability of predicting the travelers' adaptive reactions. In other words, a better understanding of the formation of cognitive maps leads to a more accurate prediction of traveler behavior.

Understanding the process of acquisition or modification of new and already acquired knowledge, skills, or preferences and modeling the travelers' cognition and learning plays an important role in traffic analysis. It also allows for the application of strategies to organize activity-based traffic control configuration to reduce network-level congestion. Employing behavioral models in transportation planning stipulates the implementation of a new generation of traffic control schemes and ultimately, enhances the mobility of transportation networks.

In this paper, we study two mental processes of a traveler: (1) developing a spatiotemporal cognitive map of a road network in traveler's mind, and (2) predictions of their travel time using information obtained from latent learning. In the first step, we focus on the global spatiotemporal image of a road network altered by changing travelers' experiences. This mental image takes shape in the memory of the traveler through the collection of information of the trips. This leads to travelers' global conceptual understanding of congestion or the degree of crowding of the network. Therefore, we try to answer how travelers draw an overall image of travel time for a transportation network and how their cognitive map is shaped up.

Furthermore, this study investigates how latent learning from experiences of travelers shapes specific parts of the cognitive map associated with parts of the road network, even for those parts where the traveler is unfamiliar with. We seek to identify how travelers can extend their experiences in order to navigate in parts of the network where they are not familiar with. This gives rise to the question about the extent to which travelers are able to make predictions of travel time for unfamiliar links based on their knowledge of some other parts of the network.

In this paper, we present some features of the two main themes of this study – cognitive map and latent learning – by utilizing thirteen hypotheses and associated laboratory-like experiments. The remainder of the paper is organized as follows. Section 2 reviews research literature. Section 3 explains the study approach, describes the experimental design and discusses the data analysis tools. Section 4 presents the results and discuss them. Section 5 concludes the paper and suggests directions for future researches.

## Literature review

'Perception updating' or 'travel time learning', which are used interchangeably, are important behavioral aspects of traveler's mental process in the dynamic or day-to-day travel demand modeling. Several learning models are utilized to formulate travelers' adaptation mechanism. Khademi et al. (2014) provided a brief review on many of the travelers' perception updating models and indicated that empirical modeling has been attempted for only a few of them.

Research performed by Horowitz (1984) was pioneering in this realm. Horowitz's model (1984) is perhaps the most elementary method for modeling the mental process of traveler's learning and adaptation. In this model, travel cost as perceived by the traveler was considered a function of traffic volume on roads. There was no explicit reference to a learning model in this study; however, a weighted sum function was introduced to capture the composite effect of past link costs on current perceived link costs. In a similar manner, Mahmassani and Chang (1986) and Cascetta (1989) utilized this mechanism determining the perceived cost as a weighted average of costs actually incurred in previous times.

Jha et al. (1998) introduced a two-step learning model. In the first step, the updated perceived travel time is calculated as a linear combination of the last experienced travel time and schedule delays. In the second step, the aggregate perception of travel time in the mind of traveler forms as a convex combination of the perceived time and the time informed by information systems. Through a more elaborated model, Iida et al. (1992) constructed a new weighted linear model for mental travel time prediction. They focused on the idea that the travel time learning model is sensitive to the accuracy of travel time predictions of previous trips, and that recent experiences have substantially greater effects on travel time prediction than older ones.

Nakayama et al.'s (1999) study created a milestone in travel time learning models. They distinguished between passenger perception and learning. The perception model was based on three assumptions: (i) perceived travel times can be expressed as weighted averages of travel times actually experienced in previous days, (ii) earlier travel times are easier to be recalled from memory, and (iii) the traveler knows the maximum and minimum travel times on each route. Their learning model, based on genetic programming, guarantees that learning inevitably improves the accuracy of predictions. In this line of research, Nakayama and Kitamura (2000) introduced a framework indicating how pieces of information are stored in the memory. Like Nakayama et al. (1999), genetic algorithms were employed on memory rules to ensure the improvement of the learning process of passengers by increasing their experiences. Nakayama et al. (2001) presented an experience-based model of travel time learning where the perceived travel time is derived from both the average of travel times experienced in latest trips as well as the difference between the longest and shortest travel times experienced. Methodologically, they employed genetic programming to ensure the convergence in the mechanism of traveler adaptation.

In another research, Oh et al. (2003) applied the well-known method of successive averages for the perception updating (learning) modeling to update the perceived travel time from the information of previous perceived travel times that ensures the convergence previously mentioned. This is in contrast with (Nakayama and Kitamura, 2000; Nakayama et al., 1999, 2001), who had used the genetic algorithm instead. The mental updates of travel time was also a simple linear function of last perceived and experienced travel times. In other words, they employed the same ideas of previous studies.

In a more elaborated representation of traveler's cognition, Arentze and Timmermans (2003) entered several cognitive aspects to the process of restoring events in traveler's memory. These include reward and its weight, and irretrievability. Reward is the intrinsic desirability of an event and its weight is the strength of association between the event and reward received. Irretrievability is a parameter capturing the decay of event over time. Following these efforts, Ettema et al. (2005) presented a model in which traveler's perception is derived from daily morning commute trips called events and represented closely the same as (Arentze and Timmermans, 2003) with the identical items (i.e. reward, weight of a reward, and irretrievability). As mentioned before, 'irretrievability' indicates the ease with which an event can be retrieved from memory, denoted as memory strength. Ettema et al. (2005) introduced the two poles of memory strength concepts: 'representativeness', which refers to the familiarity and reasonableness of event, and 'recency', which refers to the age of the event.

In an application-oriented approach, Jotisankasa and Polak (2006) suggested a day-today framework for both route choice and departure time choice. The travel time perception updating process used a linear model of the last two perceived travel times and the difference between experienced and perceived travel times of the last experience. Theoretically, their learning model does not convey a significant contribution to the traveler perception updating models. Chen and Mahmassani (2004) proposed a new framework for travel time perception updating. From the behavioral point of view, they proposed the new assumption that individuals perceive their experienced travel times with errors due either to uncertainty of travel times in memory. In their model, this stochastic error is captured as a normally distributed parameter. Their method was new because of the use of a Bayesian approach to update the learning parameters.

To further explore the learning models, Bogers et al. (2007) constructed a learning framework based on Horowitz's (1984) weighted-average approach for modeling travel time (cost) perception of previous experiences. Oh et al. (2003) utilized successive averages to update the perceived travel time from the information of previous perceived travel times. Their work was somewhat of an application of previous methods in the field of transportation.

In addition, there are studies (e.g. Ben-Elia and Shiftan, 2010; Jia et al., 2011) tried to apply the existing travel time learning model in a new era of transportation demand modeling. Khademi et al. (2014) extended the concepts of fuzzy learning model and fuzzy inference system to modeling travelers learning and adaptation. Following (Ettema et al., 2005), they brought the concept of 'representativeness' and 'recency' into a fuzzy logic based learning environment.

One of the latest studies in this domain was done by Tang et al. (2017), who developed an instance-based learning model to explain decision making in dynamic situations, where travelers make repeated choices attempting to maximize gains over the long run. In their model, an experiment is stored in the (declarative) memory of the traveler, and in line with Ettema et al.'s recency concept (2005), its activation decays over time following a power law. The most distinctive aspect of their model is that given a past experience of a segment (on a chosen path), the traveler scales up the experience for the whole segment if only a part of the segment is experienced, e.g., some type of generalization. Subsequently, Guevara et al. (2017) employed Tang et al. (2017) model in their study framework.

In brief, the development of travel time learning models owes mostly to the day-to-day demand analysis; categorized into route choice models (e.g. see, from above, Guevara et al., 2017; Khademi et al., 2014; Tang et al., 2017), activity-travel modeling within time steps (Arentze and Timmermans, 2003), day-to-day departure time choice models (Ettema et al., 2005), and combined day-to-day route and departure time choice models (Jotisankasa and Polak, 2006). All of these models attempt to construct human mental rules with mathematical relationships that can implement a learning process as well as the process of learning reinforcement. These studies provide a highly generalized model of a traveler as an economic agent, who can decide, experience, learn, and re-decide. All of these modeling tools have tried more or less to recognize the key elements of traveler learning and cognition.

As known in this detailed review, travelers learning models have been gradually improved over time. This evolution can be studied from two perspectives:

(1) Improving the framework of learning models comprising behavioral components, and
(2) Entering better components of human cognition into the learning models.

The structure of the model should be able to hold the model components together and guarantee the day-to-day convergence. On the other hand, the components of the model must be able to capture the behavioral and cognitive elements. In this paper, we examine the traveler's cognition and disregard the model structure. Here, we examine the main component in making the learning process, *cognition*. To our best knowledge and based on this detailed review, the subject of 'traveler's cognition' has not yet been developed significantly. Cognition owns a central role in traveler's learning which is the set of all mental abilities and processes related to the structure and process of knowledge: its acquisition, storage, retrieval, and manipulation, which are used by human beings, animals, and machines (Barkowsky et al., 2007). One of the many important steps in the modeling of travelers' learning is to examine the evolution of cognitive maps in their minds. Despite the high potential for practical applications of the cognitive map in transportation planning, the era of travel time modeling and perception updating lacks it. This is the first motivation of this study.

'Cognitive' denotes the dynamic processes and 'map' refers to a static image of the real environment. Cognitive map is the knowledge of physical places and cognitive mapping contains principles for establishing spatial relations among such physical places (Golledge, 1999). A cognitive map of a transportation network is a mental representation of locations tagged with time or intervals. The concept of cognitive map was first introduced by Tolman (1948). He examined two groups of rats in which the first group had no experience of exploring a maze while the other group had been placed in the maze to freely explore for a certain time. Tolman tested these two groups by placing a target (food) in the maze. He observed that the rats of group 2, who had explored the maze earlier, could find the food much faster than those of group 1. Based on this experiment, he named the sketch of the maze memorized by the rats as a "cognitive map." He concluded that when the rats explored the maze even in the absence of a specific objective (like finding the food or the exit door), the overall abstract image of the paths of the maze was projected in their memory.

Cognitive map was sporadically introduced in transportation science – such as the impact of schematic transit maps on passengers' travel decisions (Guo, 2011), the effects of travel mode on cognitive map (Mondschein et al., 2010), impacts of transportation policies on the environment (Ülengin et al., 2010), and creating a decision support model to choose the most appropriate alternative for a water crossing infrastructure (Ülengin and Topcu, 1997). However, these studies have not scrutinized the travelers' experienced-based spatiotemporal mental image. We try to bridge this gap in this study and to pave the way for finding the association between the cognitive map and navigation and the effect of vehicle navigation systems on the formation of the cognitive map. The main purpose of the cognitive map is to enable travelers to make choices related to the spatial environment. This map includes spatiotemporal information about the network elements, including route characteristics, travel time spans, distances, directions, etc. Like other mental processes, cognitive map develops over time with increasing the number of travel experiences. It can become a more effective way of actively navigating the travelers in the transportation environment. In addition to traveler navigation system aid that can be employed for finding a route, travelers can also leverage their cognitive maps as the travel guidance tool.

A second question may arise here as to whether learning occurs only when we get an answer directly from each stimulus. In his experiment on rats, Tolman (1948) found that the rats who acquired a cognitive map of the maze could retrieve the spatial properties of the maze even when the origin and the destination were different from before. He called this *latent learning*, which is learning that takes place even when there is no specific training to aim or avoid a specific consequence such as food or danger. This means that in some cases, human and animal episodic memory unconsciously records some events even when there is no specific purpose to experience an event (Gluck et al., 2008). In other words, latent learning is a form of learning that is not immediately expressed in an overt response; it occurs without any obvious reinforcement of the behavior or associations that are learned (Tavris and Wade, 1997). This is the second focus of this paper.

**Methods**

In this study, we focus on the formation of a spatiotemporal cognitive map of a network based on the factors that influence the information and recognizability of the transportation network. Moreover, the latent learning of the transportation network by travelers is characterized in this study.

*Participants*

A set of engineering undergraduate students were recruited as participants for the survey used in this study. Table 1 shows their statistical characteristics. On average, the duration of the experiment lasted around 3 hours for each participant, and the total time of data acquisition lasted one month. The sample size is tested for sufficient statistical power.

*Color-blindness test*

As the designed experiment deals with a color-coded map, participants are examined through a color-blindness test in order to mitigate the errors due to color blindness.

**Table 1** General characteristics of the participants

| 1 | Population size | 90 participants | | |
|---|---|---|---|---|
| 2 | Age (years) | Min: 20 | Max: 28 | Average: 22.71 |
| 3 | Gender | Male: 70 (77%) | | Female: 20 (23%) |
| 4 | Marital status | Single: 84 (93%) | | Married: 6 (7%) |
| 5 | Field of study | Civil engineering: 84 (93%) | | Other fields: 6 (7%) |
| 6 | Academic degree | Undergraduate: 77 (85.5%) | | Graduate: 13 (14.5%) |
| 7 | Type | University students: 90 (100%) | | |
| 8 | Driving experience | Holding driving license: 79 (88%) | | Without driving license: 11 (12%) |
| | | Driving experience of participants who have driving license (years): | | |
| | | Min: 1 | Max: 8 | Average: 3.86 |
| 9 | Color-blindness | None of the participants were diagnosed | | |

*Optimism test*

We examine the participant's natural optimism or pessimism level, as it might affect the construction of cognitive maps and the development of latent learning. The optimism test includes 48 questions about permanence, pervasiveness, hope, and personalization, through which the optimism/pessimism levels of people are determined (Seligman, 2011). According to the scores obtained from the participants in this test, their optimism level is categorized as optimistic, average or pessimistic.

*Materials of route choice experiences*

A simulation software application was specifically developed and used for data gathering. The said application lets participants choose their route and departure time and record their travel time for each link. Traffic assignment is considered the same for all participants.

The laboratory experimental procedure uses the network shown in Fig. 1 as the test environment. This hypothetical network contains 42 nodes and 67 links. In this experiment, the participants have a certain number of trips from the start point, node 1 in Fig. 1, to the finish point, node 23. During the experiment, participants are provided with the network plan (i.e. panorama route map), but are not given any prior information about the link's travel time, traffic congestion, and network environment. It is important to note that the participants are provided with the real-time and the elapsed time of the experiment in order to avoid travel time perception errors.

In the beginning of the experiment, the experimental procedure is explained to the participants with the help of a 10-minute video tutorial. It is explained that each link has a travel time of 1 to 10 minutes and there is a shortest path in which all of the links have a travel time equal to one minute. All the links are two-way and all the nodes, except for the signalized ones, have no passing or turning delay.

Participants receive the information about the number of trips that they need to make. They have the opportunity of making a tentative travel to visit various parts of the network in free flow condition, i.e., when there is little or no traffic. They are given prior information that they need to retrieve the experienced travel time after the driving task is completed. Their performance is then evaluated based on their recall accuracy of the recorded travel time.

Every signal has a pre-specified and fixed signal timing. In the experiment used in this study, it is designed in a way that it turns to red when the participants approach the junction. The participants are requested to account the signal delay time at signalized junctions for the travel time of the link they will subsequently enter. For example, consider a traveler passing node 33 toward node 37 in Fig. 1. He perceives a travel time of 5 minutes and 120 seconds delay time at node 33. His travel time from node 33 to node 37 is then 7 minutes.

There exist four distinct points in the network as noticeable features (see Fig. 1). These are the links with a statue, a monument, or a building with distinctive structure along the links. We name them *salient links* where we want to investigate their effect on the degree of the recognizability of the network elements by travelers.

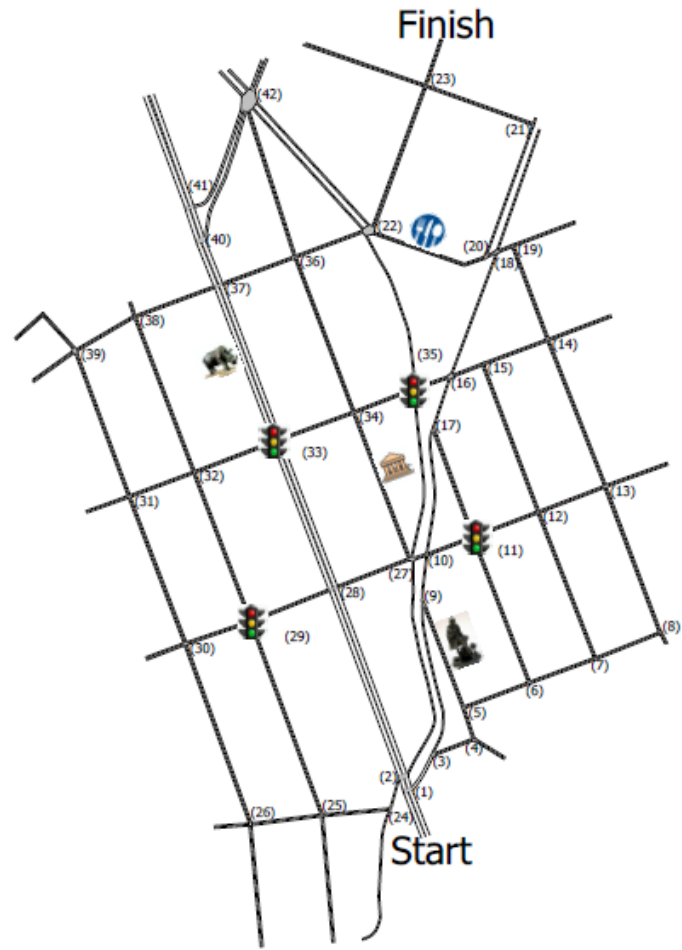

**Fig. 1.** Transportation network of the experiments.

The characteristics of the experimental data collection are shown in Table 2. As provided in this table, the network is tightly regulated to create two types of *traffic flow patterns* (TFP). In TFP 1, in general, there is less diversity among the travel times of the links leading to the nodes, i.e., the links surrounding a node differ slightly in terms of travel time. Conversely, in TFP 2, the links' travel times around the nodes differ significantly. For example, in our network presented in Fig. 1, in TFP 1, travel times of links (13→14) and (14→15) are 5 and 6 minutes, respectively, while in TFP 2, these amounts are respectively 10 minutes and 1 minute. TFP 1 and TFP 2 are independent of each other and they have been created hypothetically.

*Procedure*

According to Table 2, in this experiment, the participants (with the size of 90) are divided equally into three groups. Participants of the first, second, and third group respectively make 2, 4, and 6 trips from start to finish points in both TFP 1 and TFP 2. The indices of the nodes and links passed by the participants are recorded. Participants are informed that the two TFPs are independent. In all three groups, half of the participants start the experiment with TFP 1 and the others with TFP 2.

The links and nodes of the experiment network are divided into two groups: *tested* and *untested* nodes/links. The tested nodes/links are those that the driver has passed through at least one time, and the untested nodes/links have not been previously experienced by the driver. During the repetition of the experiment, some of the nodes/links are moved from the untested group to the tested one as the drivers broaden their experience.

At the beginning of the experiment, the participants are free to choose any path from the start point to the finish point. However, we regulate that for the next iterations, travelers are not allowed to choose the tested nodes. However, when there is no option available except the previously selected (tested) nodes, they are obliged to choose among the subset of the next node(s) having at least one untested link connected to them, if any untested link exists. Hence, the participants are implicitly forced to experience more paths and, consequently, acquire a broader knowledge of the network.

Note that taking some of the possible route-choice options can affect the formation of the cognitive map. However, we do not enforce the travelers to choose a specific route to make their trips (referred to as passive driving). The simulated transportation network is large enough to suggest multiple routes for a given origin-destination pair.

**Table 2** Experiments conducted by different groups of participants

| Group index | Number of drives from start to finish | TFP* of the Network |
|---|---|---|
| Group 1 (30 participants) | 2 | Both TFP 1 & TFP 2 |
| Group 2 (30 participants) | 4 | Both TFP 1 & TFP 2 |
| Group 3 (30 participants) | 6 | Both TFP 1 & TFP 2 |

TFP*: Traffic flow pattern

*Construction of the spatiotemporal cognitive map*

Studies in the literature concerned with the memory processing of intervals and psychological time studies can be categorized into the two following groups (Finnerty, 2015):

a. The methods work based on the actual experience of time intervals by participants. In this category of experiments, a time interval (or some time intervals) is presented to a participant through a continuous sound or flash of light. Based on the question the traveler is asked, the following methods are formed (Grondin, 2010):

  – *Verbal estimation*: The participant is asked to give a verbal estimate of the time duration in seconds, minutes, etc.

  – *Reproduction*: The participant is asked to reproduce the length of the interval by some operation.

  – *Method of comparison*: The participant has to judge the relative duration of intervals presented successively.

b. The methods that give the participants a specified amount in temporal units for an interval: The participant produces numbers, for example, by pushing a button for a duration that is judged equivalent to the target interval. This method is called *production*.

This study designs a method named "colored-visual estimation," where the participants are asked to give a visual estimate of the travel time duration in color intensities. After completion of the total number of drives under each TFP, the participants are invited to recall the perceived travel times, but not as absolute or relative values. They are asked to leverage a color-coded range shown in Fig. 2. Thus, the information of the travel time cognition of the participants is collected in the form of a color range. It is assumed that it is easier to elicit the participants' mental representation of time through it.

In the color-coded range of Fig. 2, the darker colors indicate longer travel times and the lighter colors indicate shorter ones. The discrete color range is constructed by *Adobe Photoshop CS6* by equally dividing the spectrum of red, green, and blue (RGB) color components along discrete indicator color (i.e. color IDs). In the RGB spectrum, each color in the world is obtained from the combination of these three colors. The value of each of the three components can range from 0 to 255 and each color owns a specific amount of RGB colors. Fig. 3a and Fig. 3b shows the real travel times under TFP 1 and TFP 2, respectively. The locations of signalized intersections are specified by blue boxes.

| Color ID | (1) | (2) | (3) | (4) | (5) | (6) | (7) | (8) | (9) | (10) |
|---|---|---|---|---|---|---|---|---|---|---|
| Red | 255 | 255 | 255 | 255 | 250 | 200 | 150 | 100 | 50 | 0 |
| Green | 195 | 145 | 95 | 45 | 0 | 0 | 0 | 0 | 0 | 0 |
| Blue | 195 | 145 | 95 | 45 | 0 | 0 | 0 | 0 | 0 | 0 |

**Fig. 2.** Component characteristics of the color-coded range used as the mental representation of travel time.

*Retrieval of the link-based spatiotemporal cognitive map from participant's memory*

In this type of cognitive map formation, the participants are invited to assign a color to each traveled link from their last experience to the first one. The reason for this order is to minimize the error due to memory loss as much as possible, as older events are posited to have lower memory strength. Fig. 3c depicts a sample of the spatiotemporal cognitive map from a participant who traveled only 2 times under TFP 1. The travelers are not informed of any correspondences between the range of travel time in the network and the color IDs in Fig. 2. For instance, the traveler who experiences 8 minutes as his longest travel time may assign the darkest color (ID 10) to it.

*Retrieval of the route-based spatiotemporal cognitive map from participant's memory*

After extracting the link-based spatiotemporal cognitive map from participants' memory, they are invited to assign a color to each experienced route (from start to finish). This means that the travelers are asked to express their feelings, comparatively, about the travel time of each route that they have experienced in each iteration. We predict that the accuracy and the performance of the respondents in these complex cognitive tasks depend on the number of links in the path and the diversity of the travel times in the links; hence, inaccuracy in the results due to the limitation of the memory span is anticipated. Fig. 3d shows a sample of the route-based spatiotemporal cognitive map of participant's memory.

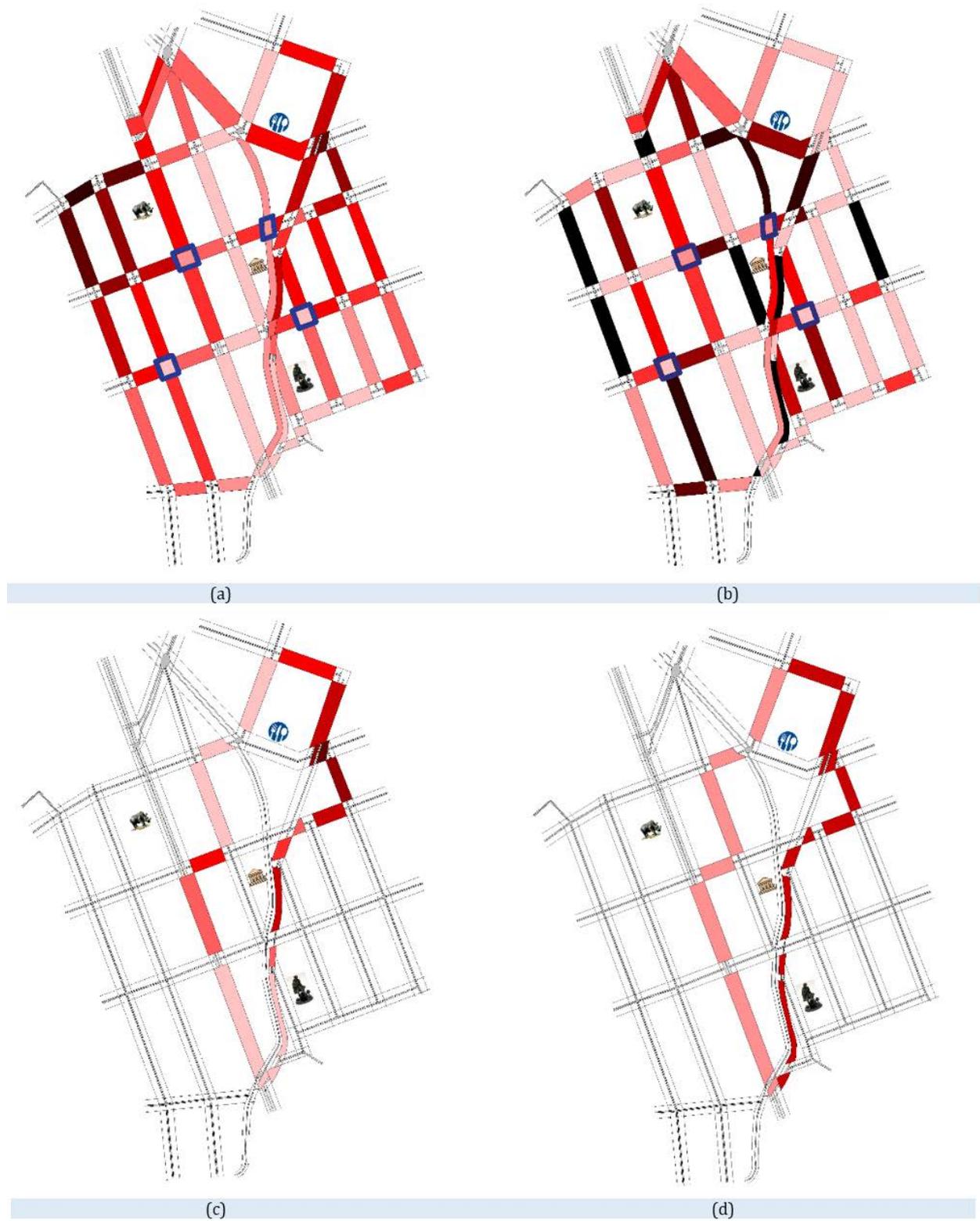

**Fig. 3. a.** Real (not recalled) travel times in TFP 1; **b.** Real (not recalled) travel times in TFP 2; **c.** A sample of link-based spatiotemporal cognitive map formed in the memory of a traveler who traveled two times under TFP 1; **d.** A sample of route-based cognitive map formed in the memory of a traveler who traveled two times under TFP 1.

*Travel time prediction using information obtained from latent learning*

In order to investigate the unconscious processes of the development of the latent learning, after the construction of the cognitive maps, the participants are asked to assign color IDs to any arbitrary set of links they think can make judgments about their possible travel time. For example, Fig. 4 shows a sample of assigning the color IDs to some untested links by a traveler whose link-based spatiotemporal cognitive map is shown in Fig. 3c. The links traveled by the participant are shown in gray color in Fig. 4 and the links for which the participant makes judgments are shown in shades of red.

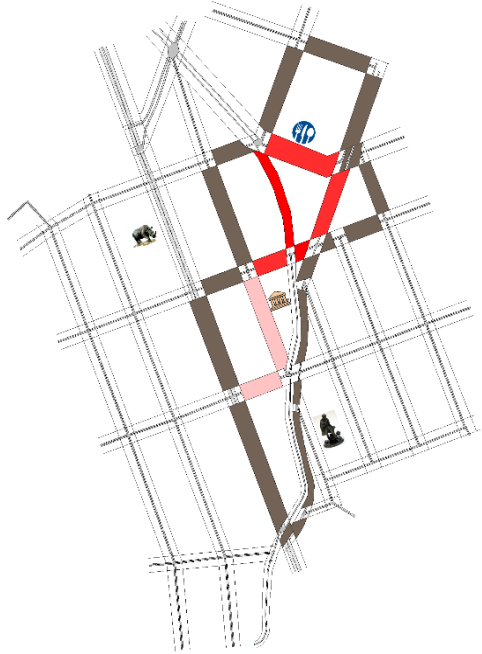

**Fig. 4.** Assigning color IDs to an arbitrary set of non-experienced links.

*Hypotheses*

*Hypotheses related to the cognitive map construction*

The first aim of this experiment is to scrutinize the construction of the spatiotemporal cognitive map of travelers. We examine the following hypotheses through the designed experiments:

**Hypothesis 1.** The link-based spatiotemporal cognitive map is affected by the uniformity or dispersion of links' travel time of the network.
**Hypothesis 2.** The route-based spatiotemporal cognitive map is affected by the uniformity or dispersion of links' travel time of the network.
**Hypothesis 3.** The cognitive map of the separate links forms a better map in travelers' memory than the cognitive map of the continuous routes.
**Hypothesis 4.** The link-based spatiotemporal cognitive map is weakened by the increase in the number of traveled links (or the number of trips).
**Hypothesis 5.** The link-based spatiotemporal cognitive map is affected by the travelers' gender.
**Hypothesis 6.** The link-based spatiotemporal cognitive map is affected by the travelers' driving experiences in other transportation networks.
**Hypothesis 7.** The link-based spatiotemporal cognitive map is affected by the presence of salient points (like statues and monuments) in the transportation network.

*Hypotheses related to the use of information obtained from latent learning*

This experiment also aims to scrutinize the area of the travelers' latent learning. Stated more precisely, it intends to investigate to what extent travelers are able to generalize their previous experiences accurately to unfamiliar networks using real-world structured record of travels. The following hypotheses are examined:

**Hypothesis 8.** The accuracy of travel time prediction is affected by the uniformity or dispersion of links' travel time of the network.
**Hypothesis 9.** The traveler's travel time prediction is affected by the number of experienced links (or the number of trips from start to finish).
**Hypothesis 10.** Travel time prediction is affected by the traveler's gender.
**Hypothesis 11.** Travel time prediction is affected by driving license holding.
**Hypothesis 12.** Travel time prediction is affected by the presence of traffic signals. In other words, unpleasant effects of signals are generalized to adjacent links.
**Hypothesis 13.** Travel time prediction is affected by the traveler's optimism level.

*Indexes to evaluate the hypotheses*

To test Hypotheses 1 to 13, it is first necessary to define the notations and indices.

*Sets and sub-sets notations*
Five sets are defined as

| | |
|---|---|
| $i$ | Index of participant, $i = 1, \ldots, 90$ |
| $L_i^E$ | Set of links that participant $i$ has already taken and obtained experienced about |
| $L_i^N$ | Set of non-experienced links that participant $i$ has not taken and not passed through |
| $R_i^E$ | Set of routes that participant $i$ has already taken and obtained experienced about |
| $\dot{L}_i^E$ | Set of links containing salient features that participant $i$ has already taken and obtained experienced about |
| $\ddot{L}_i^E$ | Set of signalized links that participant $i$ has already taken and obtained experience about (a signalized link is link at the beginning of which there is a traffic light) |
| $S_i^E$ | Set of links that participant $i$ chooses to estimate their travel times (without having the experience of passing them) |

*Traffic characteristic index*
For each TFP 1 and TFP 2, we define the travel time diversity index as follows:

$$K^p = \frac{1}{N}\sum_{n=1}^{N} k_n^p; \qquad (1)$$

where

$$k_n^p = \frac{1}{M_n}\sum_{k=1}^{M_n}\sum_{k=1}^{M_n}|t_k^p - t_l^p| \qquad (2)$$

The notations are:

| | |
|---|---|
| $p$ | TFP index |
| $n$ | Node index ($n=1,\ldots,N$) |
| $N$ | Total number of nodes in the network |
| $M_n$ | Number of links that node $n$ receives |
| $k, l$ | Index of links |
| $t_k^p, t_l^p$ | Real (existing) travel times of link $k$ and $l$ under TFP $p$, respectively |
| $k_k^p$ | Mean of pairwise differences between travel times of the links belong to node $n$ (under TFP $p$) |

In Eq. 1, it is clear that when link travel times around a node (say $n$) are close to each other, the differences between them, and consequently $k_k^p$, are low. If $k_k^p$ are low throughout the network, $K^p$ is low which implies the uniformity of the travel times of the links throughout the network. In the experiment environment, the network is tightly regulated (as mentioned before) so as to create two fixed types of traffic flow patterns. It gives two travel time diversity indices, $K^1$ equal to 1.5 (for TFP 1) and $K^2$ equal to 4.4 (for TFP 2).

*Real travel time vs. perceived travel time*

*Real travel time.* Every signal has a pre-specified and fixed signal timing. In our experiment, it is designed in a way that it turns to red when the participants approach the junction. As we ask the participants to add the signal delay time to the travel time of the links located in front of the signal, we calculate the real travel time of link $k$ under TFP $p$ ($t_k^p$) by

$$t_k^p = \sum_{k,s} te_k^p + \tau_{k,s}^p \delta_{k,s} \qquad (3)$$

where the notations are:
- $k$      Index of link
- $p$      Index of traffic flow pattern (TFP) ($p$=1, 2)
- $s$      Index of signal
- $te_k^p$      Travel time of link $k$ under TFP $p$ without assigning the delay time of traffic signals
- $\tau_{k,s}^p$      Length of the red phase of signal $s$ under TFP $p$
- $\delta_{k,s}$      A dummy variable equal to $\begin{cases} 1 & \text{if signal } s \text{ is jointed to the begining of link } k \\ 0 & \text{otherwise} \end{cases}$

If a link does not contain a signal ahead of it, then $t_k^p = te_k^p$.

*Perceived travel time.* As mentioned before, in this study, the participants are invited to recall the travel times, but not as absolute or relative values. We ask them to leverage a color-coded range shown in Fig. 2. Each color assigned by a participant has a code which indicates the travel time in minutes. The extracted maps are converted to numbers using the codes in Fig.2 and are used as inputs for calculating the perceived travel time of link $k$ by participant $i$ under TFP ($\hat{t}_{k,i}^p$):

$$\hat{t}_{k,i}^p = tr_{k,i}^p \frac{m_i^p}{10} \qquad (4)$$

where the notations are
- $k$      Index of link
- $i$      Index of participant
- $p$      Index of traffic flow pattern
- $tr_{k,i}^p$      Recalled travel time under TFP $p$ by participant $i$; it is the index of the color assigned to link $k$ (from Fig. 2)
- $m_i^p$      Maximum experienced travel time by participant $i$ under TFP $p$ ($m_i^p = \max te_k^p, \forall k \in L_i^E$)

As mentioned before, each link of the network has a travel time from 1 to 10 minutes, respectively the theoretical minimum and maximum travel times in both TFPs. The traveler may never experience a 10-minute travel time. S/He must assign colors to experienced travel times from the color-coded range shown in Fig. 2. They are required to assign the darkest color to the highest experienced travel time and the lightest color to the shortest one in a relative comparison context. Therefore, in order to make an absolute comparison between the real (existing) travel times and the recalled travel times, we leverage $\frac{m_i^p}{10}$ for scaling the recalled travel times using the highest real travel time in the network (10 minutes). It is important to note that it is not necessary to use the scaling variable for the least amount of travel time (1 minute), as all the travelers definitely experience this travel time in the start link. That is, both the starting links (1→2 and 1→3 in Fig.1) have 1-minute travel time and each participant invariably experiences one of them.

*Indices related to participants responses*

*Index 1.* To evaluate the retrievability of participant $i$ in the construction of the link-based spatiotemporal cognitive map under TFP $p$, the following index is designed:

$$PRAL_i^p = \frac{1}{|L_i^E|} \sum_{L_i^E} |t_k^p - \hat{t}_{k,i}^p| \qquad (5)$$

where we have
- $i$      Index of participant
- $p$      Index of traffic flow pattern (TFP)
- $k$      Index of link
- $|L_i^E|$      Number of elements of set $L_i^E$ (i.e. total number of the experienced links by participant $i$)
- $t_k^p$      Real travel time (from Eq. 3)

$\hat{t}_{k,i}^p$     Perceived travel time (from Eq. 4)

This index calculates the average of the differences between the real (existing) travel times and the perceived travel times of the links over the set of experienced links ($L_i^E$), by participant $i$ under TFP $p$. The larger the value of $PRAL_i^p$, the lower the accuracy of participant's in the construction of the link-based spatiotemporal cognitive map.

*Index 2.* To evaluate the retrieval ability of participant $i$ in the construction of the route-based spatiotemporal cognitive map under TFP $p$, the following index is designed:

$$PRAR_i^p = \frac{1}{|R_i^E|}\Sigma_{R_i^E}\left|r_z^p - \hat{r}_{z,i}^p \frac{m_i^p}{10}\right| \tag{6}$$
$$\text{in which } r_z^p = \frac{1}{|L_i^E|}\Sigma_{L_i^E} t_k^p$$

where the notations are:
- $i$     Index of participant
- $p$     Index of traffic flow pattern (TFP)
- $k$     Index of link
- $z$     Index of route
- $|L_i^E|$     Number of elements of set $L_i^E$ (i.e. total number of the links experienced by participant $i$ when he passes the route)
- $|R_i^E|$     Number of elements of set $R_i^E$ (i.e. total number of the routes experienced by participant $i$) where $|R_i^E| = 2, 4, or\ 6$
- $t_k^p$     Real (existing) travel time of link $k$ under TFP $p$ from Eq. 3
- $\hat{r}_z^p$     Recalled travel time of route $z$ under TFP $p$
- $r_z^p$     Experienced travel time of route $z$ under TFP $p$
- $m_i^p$     Maximum experienced travel time by participant $i$ under TFP $p$ ($m_i^p = \max te_k^p, \forall k \in L_i^E$, where $te_k^p$ is the travel time of link $k$ under TFP $p$ without assigning the delay time of traffic signals)

This index calculates the average of the differences between the existing traffic condition over the set of experienced routes ($\frac{1}{|L_i^E|}\Sigma_{L_i^E} t_k^p$) and passenger $i$'s recalled sense of traffic condition ($\hat{r}_{z,i}^p \frac{m_i^p}{10}$) under TFP $p$. Similar to above, a large value of $PRAR_i^p$ corresponds to a lower accuracy of participants in the construction of the route-based spatiotemporal cognitive map.

*Index 3.* To evaluate the travel time prediction ability of participant $i$ in the process of the development of his latent learning, the following index is designed:

$$TTPA_i^p = \frac{1}{|S_i^E|}\Sigma_{S_i^E}(10 - |t_k^p - \hat{t}_{k,i}^p|) \tag{7}$$

where the notations are:
- $i$     Index of participant
- $p$     Index of traffic flow pattern (TFP)
- $k$     Index of link
- $S_i^E$     Set of links that participant $i$ chooses to estimate their travel times (without having the experience of passing them)
- $|S_i^E|$     Number of elements of set $S_i^E$
- $t_k^p$     Real travel time (from Eq. 3)
- $\hat{t}_{k,i}^p$     Perceived travel time (from Eq. 4)

This index rates the accuracy of the participants' answers by calculating the average sum of the differences between the real (existing) travel time and the estimated travel time of links over the set of estimated links ($S_i^E$) by participant $i$ under TFP $p$. $S_i^E$ is the set of links that the participant $i$ chooses to estimate their travel times without having the experience of passing them. If a participant estimates an exact travel time for a link, he gains 10 points for his estimation while with an increase in the difference between the real and estimated travel time, this score decreases. For example, if the assignment of color-IDs to the travel time of a link has a 2-point difference from the real travel time, the participant gains (10-2)/1=8 points. If the estimation on 2 links has 1 point deviation for each link, he gains [(10-1)+(10-1)]/2=9 points. Larger values of $TTPA_i^p$ indicate a higher ability of travel time prediction by participants.

*Index 4.* In order to investigate the effect of the salient features (i.e. statues, monuments, or buildings with distinctive structures) on the ability of participant $i$ in the construction of the link-based spatiotemporal cognitive map, the following index is defined:

$$SE_i^p = \frac{1}{|\dot{L}_i^E|} \sum_{\dot{L}_i^E} |t_k^p - \hat{t}_{k,i}^p| \tag{8}$$

where:
$i$      Index of participant
$p$      Index of traffic flow pattern (TFP)
$k$      Index of link
$\dot{L}_i^E$      Total number of experienced salient links by participant $i$ (salient links are the links having salient or noticeable features)
$|\dot{L}_i^E|$      Number of elements of set $\dot{L}_i^E$
$t_k^p$      Real travel time (from Eq. 3)
$\hat{t}_{k,i}^p$      Perceived travel time (from Eq. 4)

This index calculates the average of the differences between the real (existing) travel time and recalled travel time of links with salient objects along the street for participant $i$ under TFP $p$. If a participant has not traveled in these links, this index is not calculated for him. It is obvious that a larger value of $SE_i^p$ indicates a lesser positive effect of the presence of the salient objects on participant's link-based spatiotemporal cognitive map.

*Index 5.* In order to investigate the signalized intersection effect on the ability of participant $i$ in the construction of the link-based spatiotemporal cognitive map, the following index is defined:

$$SIE_i^p = \frac{1}{|\dot{S}_i^E|} \sum_{\dot{L}_i^E} (10 - |t_k^p - \hat{t}_{k,i}^p|) \tag{9}$$

where the nations are:
$i$      Index of participant
$p$      Index of traffic flow pattern (TFP)
$k$      Index of link
$\dot{S}_i^E$      Set of signalized-links that participant $i$ estimates their travel time (without having the experience of passing them)
$|\dot{S}_i^E|$      Number of elements of set $\dot{S}_i^E$
$t_k^p$      Real travel time of signalized-link $k$ under TFP $p$ (from Eq. 3)
$\hat{t}_{k,i}^p$      Perceived travel time of signalized-link $k$ under TFP $p$ (from Eq. 4)

This index calculates the sum of the differences between the real (existing) travel time and the estimated travel time of signalized links over the set of estimated links, by participant $i$ under TFP $p$. If a participant estimates the exact travel time of a signalized link, he gains 10 scores for that estimation and with an increase in the difference between the real and estimated travel time, this score decreases. If a participant has not traveled through signalized links, this index is not calculated for him. The larger the values of $SIE_i^p$, the greater the positive effect of the presence of signals on participants' ability of travel time prediction.

*Index 6.* The following index is defined to measure every traveler's accuracy of the prediction of travel times of non-experienced links within the experiment of the "Travel time prediction using information obtained from latent learning":

$$TAL_i^p = \frac{\sum_{S_i^E} I[(\hat{t}_{k,i}^p - t_k^p) > 0] - \sum_{S_i^E} I[(\hat{t}_{k,i}^p - t_k^p) < 0]}{|S_i^E|} \quad (10)$$

where $k$ is the link index, $t_k^p$ and $\hat{t}_{k,i}^p$ are respectively the real (existing) and recalled travel times of link $i$ under TFP $p$, $S_i^E$ is the set of links that participant $i$ chooses to estimate their travel times without having the experience of passing them, $I[\cdot]$ is an indicator function which gives 1 if the condition in the bracket is satisfied, and zero otherwise. The first indicator function $(I[(\hat{t}_{k,i}^p - t_k^p) > 0])$ points to the links for which the traveler has perceived travel times more than the real amounts. Stated analogously, it counts the number of links for which the stated travel times is higher than the real (existing) travel times (i.e. those the participants overestimate the travel time). In the same manner, the second bracket $(I[(\hat{t}_{k,i}^p - t_k^p) < 0])$ counts the underestimation cases.

## Results and discussion

*Results*

Fig. 5 illustrates the descriptive statistics of all the six proposed indexes: *PRAL* in Eq. 5, *PRAR* in Eq. 6, *TTPA* in Eq. 7, *SE* in Eq. 8, *SIE* in Eq. 9, and *TAL* in Eq. 10. This figure represents the mean, standard deviation, median, interquartile range, skewness, and kurtosis values of the two proposed traffic flow patterns.

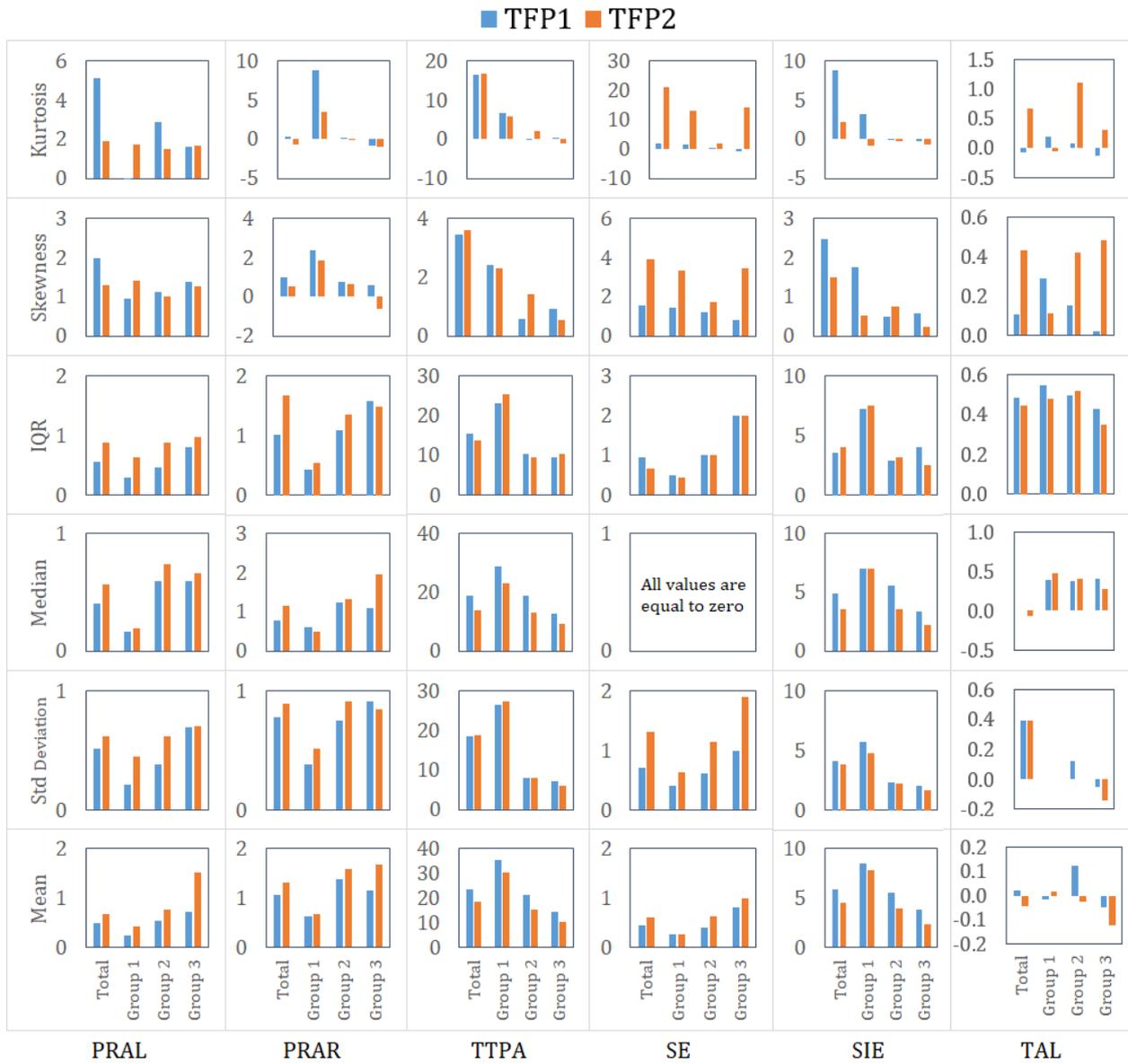

**Fig. 5.** Descriptive statistics of indexes 1-6 with respect to different groups in Table 2 and for two traffic flow patterns (TFP).

As investigated by Kolmogorov-Smirnov goodness-of-fit test, all of the proposed indexes in the previous section are non-normally distributed. Therefore, to evaluate the hypotheses proposed in this research, nonparametric statistical tests are employed. Table 3 shows the results of the statistical tests related to Hypotheses 1 to 7. Similarly, Table 4 shows the results of the test regarding Hypotheses 8 to 13. The 3rd column of the tables points to the nonparametric tests utilized.

It should be noted that according to Table 2, the participants in this experiment (with the size of 90) are divided equally into three groups. Participants of the first, second, and third group make 2, 4, and 6 trips respectively from start to finish points of the network (Fig. 1) in both TFP 1 and TFP 2. For all the indexes (1 to 6), the subscript points to the TFP and the superscript indicates the respondent, e.g. $PRAL_i^1$ denotes the value of the $PRAL$ index under TFP 1 for respondent $i$.

**Table 3** Results of the statistical tests of Hypotheses 1 to 7

| Hypotheses | Explanation | Statistical test | What is checked by the statistical test? | Population* | Sig. Value | Verification |
|---|---|---|---|---|---|---|
| Hypothesis 1 | The link-based spatiotemporal cognitive map is affected by the uniformity or dispersion of the links' travel time of the network. | Wilcoxon matched-pairs signed-ranks test | The difference between the two sets of $\{PRAL_i^1 \| i = 1:90\}$ and $\{PRAL_i^2 \| i = 1:90\}$ is significant (see Eq. 5 for $PRAL$). | 90 | 0.006 | ✓ |
| Hypothesis 2 | The route-based spatiotemporal cognitive map is affected by the uniformity or dispersion of the links' travel time of the network. | Wilcoxon matched-pairs signed-ranks test | The difference between the two sets of $\{PRAR_i^1 \| i = 1:90\}$ and $\{PRAR_i^2 \| i = 1:90\}$ is significant (see Eq. 6 for $PRAR$). | 90 | 0.545 | ✗ |
| Hypothesis 3 | The link-based spatiotemporal cognitive map forms better than the route-based spatiotemporal cognitive map in the travelers' memory. | Wilcoxon matched-pairs signed-ranks test | The difference between the two sets of $\{PRAL_i^1 \cup PRAL_i^2 \| i = 1:90\}$ and $\{PRAR_i^1 \cup PRAR_i^2 \| i = 1:90\}$ is significant (see Eq. 5 and 6 for $PRAL$ and $PRAR$). | 180 | 0.000 | ✓ |
| Hypothesis 4 | The link-based spatiotemporal cognitive map is affected by the number of traveled links (or the number of trips). | Kruskal-Wallis one-way analysis of variance by ranks | If the 180-member sets of $\{PRAL_i^1 \cup PRAL_i^2 \| i = 1:90\}$ are divided into three subsets according to the number of iterations (i.e. trips) from start to finish, which are 2, 4, and 6, then there will be a significant difference between them (see Eq. 5 for $PRAL$). | 180 | 0.000 | ✓ |
| Hypothesis 5 | The link-based spatiotemporal cognitive map is affected by travelers' gender. | Kolmogorov-Smirnov test for two independent samples | When participants are divided into two categories by gender type (male and female), the differences between these two sets based on the retrieval ability values $(PRAL_i^1 \cup PRAL_i^2)$ are significant (see Eq. 5 for $PRAL$). | 180 | 0.880 | ✗ |
| Hypothesis 6 | The link-based spatiotemporal cognitive map is affected by travelers' driving experience in other transportation networks. | Kruskal-Wallis one-way analysis of variance by rank | When participants are divided into three categories by driving experience ((i) no driving experience, (ii) driving experience between 1 to 5 years, and (iii) driving experience more than 5 years), the differences between these sets based on the retrieval ability values $(PRAL_i^1 \cup PRAL_i^2)$ are significant (see Eq. 5 for $PRAL$). | 180 | 0.050 | ✓ |
| Hypothesis 7 | The link-based spatiotemporal cognitive map is affected by the presence of salient features (like statues and monuments) in the transportation network. | Wilcoxon matched-pairs signed-ranks test | The differences between the two sets of $\{PRAL_i^1 \cup PRAL_i^2 \| i = 1:90\}$ and $\{SE_i^1 \cup SE_i^2 \| i = 1:90\}$ are significant (see Eq. 5 and 8 for $PRAL$ and $SE$). | 180 | 0.479 | ✗ |

*For Hypotheses 3-7, the data of both TFP 1 and TFP 2 are appended, resulting in a population of 180.*

**Table 4** Results of the statistical tests of Hypotheses 8 to 13

| Hypotheses | Explanation | Statistical test | What is checked by the statistical test? | Population* | Sig. Value | Verification |
|---|---|---|---|---|---|---|
| Hypothesis 8 | The accuracy of travel time predictions is affected by the uniformity or dispersion of links' travel time of the network. | Wilcoxon matched-pairs signed-ranks test | The difference between the two sets of $\{TTPA_i^1 \mid i = 1:90\}$ and $\{TTPA_i^2 \mid i = 1:90\}$ is significant (see Eq. 7 for $TTPA$). | 90 | 0.000 | ✓ |
| Hypothesis 9 | The travelers' travel time predictions are affected by the number of the experienced links (i.e. the number of the trips from start to finish). | Kruskal-Wallis one-way analysis of variance by rank | If the 180-member set of $\{TTPA_i^1 \cup TTPA_i^2 \mid i = 1:90\}$ is divided into three subsets according to the number of iterations (i.e. trips) from start to finish, which are 2, 4, and 6, there will be a significant difference between them (see Eq. 7 for $TTPA$). | 180 | 0.047 | ✓ |
| Hypothesis 10 | The travelers' travel time predictions are affected by the travelers' gender. | Kolmogorov-Smirnov test for two independent samples | When participants are divided into two categories by gender type (male and female), the difference between these two sets based on retrieval ability values $(TTPA_i^1 \cup TTPA_i^2)$ are significant (see Eq. 7 for $TTPA$). | 180 | 0.490 | ✗ |
| Hypothesis 11 | The travelers' travel time predictions are affected by the travelers' driving experiences in other transportation networks. | Kruskal-Wallis one-way analysis of variance by rank | When participants are divided into three categories by driving experience ((i) no driving experience, (ii) driving experience between 1 to 5 years, and (iii) driving experience more than 5 years), the differences between these sets based on retrieval ability values $(TTPA_i^1 \cup TTPA_i^2)$ are significant (see Eq. 7 for $TTPA$). | 180 | 0.050 | ✓ |
| Hypothesis 12 | The travelers' travel time predictions are affected by the presence of traffic signals. To express it more clearly, unpleasant effects of signals are generalized to adjacent links. | Wilcoxon matched-pairs signed-ranks test | The difference between the two sets of $\{TTPA_i^1 \cup TTPA_i^2 \mid i = 1:90\}$ and $\{SIE_i^1 \cup SIE_i^2 \mid i = 1:90\}$ is significant (see Eq. 7 for $TTPA$ and Eq. 9 for $SIE$). | 180 | 0.000 | ✓ |
| Hypothesis 13 | The travelers' travel time predictions are affected by the travelers' optimism level. | Spearman bivariate correlation | When participants are divided into three categories by the optimism/pessimism type ((i) optimistic traveler, (ii) average people, and (iii) pessimistic traveler), the differences between these groups are significant. In other words, there is a significant correlation between the participants' optimism/pessimism level and the set of $TAL_i$ (see Eq. 10 for $TAL$). | 180 | 0.031 | ✓ |

*For Hypotheses 9-13, the data of both TFP 1 and TFP 2 are appended, resulting in a population of 180.*

*Discussion*

For indexes 1 (*PRAL* in Eq. 5), 2 (*PRAR* in Eq. 6), 4 (*SE* in Eq. 8), and 6 (*TAL* in Eq. 10), lower values of the index correspond to better responses. By contrast, in the other indexes (index 3 (*TTPA* in Eq. 7) and 5 (*SIE* in Eq. 9)), higher values correspond to better responses. Stated analogously, the higher values of *PRAL* (in Eq. 5), *PRAR* (in Eq. 6), and *SE* (in Eq. 8) implies the lower accuracy of the participant's cognitive map construction; on the contrary, the lower amounts for *TTPA* (in Eq. 7) and *SIE* (in Eq. 9) indicates the lower accuracy of the participant's latent learning extension. Taking into account this phenomenon, almost in all cases in Fig. 5, we observe that the precision of the response for TFP 1 is greater than TFP 2 (considering both mean and median values).

The values of the standard deviation and interquartile ranges in Fig. 5 indicate that almost for all the indexes and over all the categories of participants, for TFP 2, the amounts are spread out over a wider range of values for TFP 2 compared to TFP 1. Except for only one case (Group 3 – TFP 2 for *PRAR*), the skewness is positive (right-skew) for all the indexes, pointing to the fact that the mean is greater than the median in almost all cases. The kurtosis value in some cases is very high which indicates the heavy-tailed distributions. But overall, no specific pattern can be observed for kurtosis values among different cases.

Regarding the statistical test on the hypotheses, our main achievements are as follows:

- From Hypothesis 1 (in Table 3), the results of the statistical tests show that the accuracy of the travelers in the construction of the link-based spatiotemporal cognitive map is affected by the dispersion of the links' travel time in the transportation network. The higher the variance of the travel times in the network, the more the degree of the errors in the retrieval of travel times. This evidence is consistent with the payoff variability effect proposed by Erev and Barron (2005). They showed that the high payoff variability results in moving choice behavior to randomness. In fact, in our experiment when the participants encountered multiple choices in terms of travel time, their decisions were blended with randomness. This observation is also generalized to the development of the latent learning to untested parts of the network, according to Hypothesis 8, supported in Table 4. In other words, the uniformity of travel time throughout the network lightens the cognitive burden, on either the construction of the spatiotemporal map or the generalization of the map to the non-experienced parts using latent learning.

- The above finding is not observed in the route-based cognitive map construction (see Hypothesis 2 in Table 3). There is no significant pattern or noticeable difference of Index 2 between TFPs.

- In general, the result of the test on Hypothesis 3 (in Table 3) shows that the link-based spatiotemporal cognitive map of the travelers is different from the route-based cognitive map. This may be because of working memory limitation that affects the arithmetic ability of respondents. These results are consistent with the fact that working memory load has its impact on information retrieval from memory (Anderson et al., 1996).

- According to Hypothesis 4, the consistency of the spatiotemporal cognitive map with reality is affected by the number of the traveled links (or the number of the trips). It means that continuous replication and repeated traveling in a path reinforce the image and the data of the path in the mind of the traveler, similar to what happens to professional drivers. This is reflected in Fig. 5 by the increase in the values of *PRAL* (see the mean values of *PRAL* in the last row and the first column) with the increase in the number of trips in the network (from Group 1 to 3). As expected intuitively, it indicates that the information about the unchangeable traffic pattern of a network is reinforced when the experiment is repeated multiple times. In fact, by increasing the frequency, the level of familiarity of travelers is increased (Rogers et al., 2015). The idea that more experiences help more in the retrieval of the data is not a new idea and it is only an example of data storage in the memory through the so-called human reinforcement learning mechanism.

- In these experiments, travelers' gender is not the determinant of the mental map construction; having driving experience is on the borderline. In other words, neither the link-based spatiotemporal cognitive map (Hypothesis 5 in Table 3) nor the travelers' travel time predictions in non-experienced links (Hypothesis 10 in Table 4) are affected by the travelers' gender. One explanation for this outcome could be the effects of the limited number of participants in our study. The effect of driving experiences on the link-based spatiotemporal cognitive map (Hypothesis 6 in Table 3) and travelers' prediction ability on latent learning (Hypothesis 11 in Table 4) are significant. It means that the travelers who already have driving experience could better recall the travel time and generalize the travel time to non-experienced links after considering the travel time of the experienced links.

- For a given standardized test, there are usually one or more of the following levels of significance reported: 68%, 85%, 90%, 95%, and 99% (Kaufman and Lichtenberger, 2005). In some fields of study, based on the magnitude of the p-value, different terms are ascribed to the significance of test as: 'unacceptable', 'fair', 'good', or 'excellent' (Cicchetti, 1994). With each of these degrees of significance level, depending on the context of the problem, different views can be expressed. In general, the significance level in psychosocial and medical studies is selected with extra care because the error in these studies directly affects the health or life of humans. Indeed, in psychological and medical studies (or sciences like aerospace), researchers will do their best to eliminate the probability of type I error of the inferences. The significance level regarding Hypothesis 6 and 11 with $\alpha = 0.05$ is borderline. In fact, we reject the null hypothesis at $\alpha = 0.01$; however, we do not reject it at $\alpha = 0.05$ (although there is still some doubt about it).

- In our study, the presence of the salient features (like statues and monuments) in the network does not affect the cognitive map. This means that according to our experiment, there is no difference between recalling the travel times of the links with adjacent salient features and other ones (Hypothesis 7 in Table 3) in the specific experimental context of this analysis. At first, it may seem counterintuitive that for individuals who do not have complete knowledge of the road network arrangement, the memory of cities is shaped around 'anchors', mainly salient features in urban space, around which subjective knowledge is expanded and recalled (Manley et al., 2015). However, the current experiment indicates that in a new environment with new events, where the drivers receive information and decide based on perceptual and conceptual processes of the short-term memory, salient features do not have a significant impact on the route choices. It is because of the fact that (1) according to (Potter, 2012), in the conceptual short-term memory, most cognitive processing occurs without review or rehearsal of material in standard working memory and with little (or no conscious) reasoning, and (2) in a new environment, the entire routes and network components are so different and new to the driver that a salient feature cannot be considered as an outstanding image producing a meaningful effect.

- In this experiment, the participants make trips several times in a network they have not previously experienced in real life. They are obliged to take different routes between two fixed points. The physical characteristics of the routes, the surrounding environment, and their urban elements are nearly the same. The *TTPA* (latent learning score term) decreases by increasing the number of iterations (see mean values of *TTPA* in Fig. 5 at the intersection of the last row and the third column). The same happens for *SIE*, latent learning score term, in dealing with signalized intersections (see mean values of *SIE* in Fig. 5 at the intersection of the last row and the fifth column). This means that having a general (latent) knowledge of the traffic conditions in transportation networks from the past did not help the travel in making a better prediction (for non-experienced travel time) when he repeated the experiences, yet through different routes and confronting scattered information from various dissimilar experiences. In other words, predicting the travel time for non-experienced links becomes more difficult when the participants increase the experience of very similar events. This phenomenon may be coherent with the proposed idea of Gluck et al. (2008) that when we park our cars in the same large parking lot every day, we may confuse the episodic memories of all the prior highly similar parking events, making it difficult to remember exactly where the car was parked that day. This should not be confused with the idea that repeated experiences on one or a few routes could reinforce the information of the routes in the memory. The example of Gluck et al. (2008) conforms to the idea of Lowry (2014) which stated that, generally, repeated exposure strengthen semantic memory. By contrast, repeated exposure to very similar events may be able to weaken episodic memory for any one event. However, contrary to Lowry (2014) and Gluck et al. (2008), in this paper, we do not observe this phenomenon on the quality of remembering the travel time of experienced links (where the repartition reinforces the information in the memory), but it emerges when the participants want to guess the travel time of the non-experienced links using their latent learning. Stated more precisely, repeated experiences of very similar elements of the network may be able to weaken the precision of the development of the spatiotemporal cognitive map of the network to non-experienced parts.

− The presence of the traffic signals results in the differences between the latent learning of the signalized links and other links' travel times. This is due to the strong influence of negative feelings from delays at traffic signals on travelers.

− It is observed that despite all our previous experience and knowledge of trips and the transportation system, the manner leveraged to use the latent learning depends on our level of optimism and pessimism. In other words, the optimism and pessimism level of the people forms the image of the non-experienced parts of the network from the experienced ones (Hypothesis 13 in Table 4). Table 5 reports the mean value of *TAL* in Eq. 10 (travel time prediction error) for three categories of participants in terms of optimism level and with respect to the two aforementioned traffic flow patterns. The results point to the idea that the normal travelers have a better prediction of travel time for non-experienced parts of the network compared to the highly optimistic and pessimistic ones. According to this experiment, optimistic people exhibit judgment bias toward optimistic forecasts and vice versa for pessimistic people.

**Table 5** Travel time prediction error for different optimism levels (derived from Eq. 10)

| Category | $\overline{TAL}$ |
|:---:|:---:|
| **Optimistic driver** | 0.06 |
| **Normal driver** | 0 |
| **Pessimistic driver** | -0.04 |

- Our observations on the experiments reveal that, for making a judgment about possible travel time of non-experienced links, the participants are mostly inclined toward those sets of links closed or joined to the experienced ones.

## Conclusion

This paper focuses on the concept of cognitive map of the transportation network, which is the traveler's global conceptual understanding of the congestion or the degree of crowding of the network. It represents the travelers' internal image of experienced or non-experienced but predictable travel time, the degree of congestion, or the level of reliability of the network. We first showed how different characteristics of travelers or different network environment contribute to the human abilities in the formation of the spatiotemporal cognitive map.

We also discussed the concept of latent learning, which points to a learning mechanism in the absence of reinforcement. The latent learning of a traveler from previous experience, when combined with the current experiences, may shape some parts of the mental image, even for those parts of which the traveler has no experience.

The question is how one could translate the outcomes of the paper into practical implications especially with the development of technology at this time. In the context of the technological advancement, a well-known management consultant, Walter Bennis, said the factory of the future will have only two employees, a dog, and a human who is only to feed the dog (Paradies and Unger, 2000). Such an exaggerated picture can be extended to the transportation environment. Like all the systems in the world, transportation systems move toward automation and expert systems. We may have a forward-thinking picture of flying taxis and suspended magnetic pods helping change the dream of a dynamic driverless world into a reality.

Nowadays, a good practice of technological development is community-driven traffic and navigation consumer software applications, such as Waze, Google Maps, and Apple Maps. They provide a graphical representation of the network, which enables a better understanding of network dynamics and the urban patterns of the travelers. In this situation, a challenging question may be raised: Now when we have the systems automatically provide us with the best route (the shortest or the least-congested), what is the need to use individual abilities to make route choices? Stated more precisely, now, why do we need to examine the process of forming a cognitive map of the transportation network in the memory of travelers? In response to this question, we do not want to mention with an unreasonable justification that despite the automation of the system, individuals must have a cognitive exercise to maintain their mental ability. Neither do we even convince ourselves with the naïve idea that such a study is necessary because in some cases the navigation system may fail and people need to use their spatiotemporal cognitive map. We do not even want to insist on the issue that the subject of the mind and cognitive map is helpful to the environment that have not yet been infiltrated by modern technologies.

In fact, in the presence of the community-driven traffic and navigation applications, although the system eliminates a lot of individual trials and errors for travel decisions, it cannot be denied that the cognitive maps of the transportation network are still forming in the drivers' minds. That is to say, the route guidance and navigation process may make individuals free of random or unreasonable decisions of pathfinding, yet still, each choice and travel form the cognitive map. How the cognitive map is formed because of the presence of such facilities is the subject of another research that follows this paper. The importance and necessity of this subject will be pronounced when we want to increase the satisfaction of users and boost their compliance rate.

Another matter of concern is, as a result of the combination of (1) information provided by the systems, (2) the data that the traveler has already had, and ultimately, (3) the knowledge that he gains from his experience, how the degree of individual confidence to the system changes and how it can be increased. Also, when a significant portion of urban citizens utilizes and follow the navigation system, the system may be able to distort the cognitive maps of individuals due to errors or data deficiencies (or even due to intentional reasons, for example, to approach to a near optimal system condition). In this case, the system forms the mind map contrary to the existing reality and draw a different perceptual picture from the real world. The consequent negative (or even positive) impacts of the cognition distortion can be an important topic for further research.

To overcome the limitations of the current study, such as employing only the university students as participants, as well as the limited number of participants with non-normal distribution of age and gender, we would try to generalize the experiment data collection in a future study. This will hopefully help to strengthen or revise our conclusions. More investigation with real or simulated data from other networks with different configurations and demand patterns are required to further explore the findings of this research. Quantifying the impacts of cognitive maps and identifying other factors contributing into the formation of them are other directions for future researches. In addition, incorporating the natural spatial and temporal learning abilities based on the findings of this research to design new generation of travelers' navigation tools opens a new avenue of studies.